\documentstyle[epsfig,axodraw,12pt]{article}
\begin{document}
\centerline{\Large \bf $k_T$ factorization of exclusive $B$ meson
decays}\vskip 0.5cm

\centerline{Hsiang-nan Li}

\centerline{Institute of Physics, Academia Sinica, Taipei, Taiwan
115, Republic of China}

\centerline{Department of Physics, National Cheng-Kung
University,}

\centerline{Tainan, Taiwan 701, Republic of China}

\vskip 1.0cm
\begin{abstract}
I review recent progress on exclusive $B$ meson decays made in the
perturbative QCD approach, concentrating on the evolution of the
$B$ meson wave function in $k_T$ factorization, radiative decays,
polarizations in $VV$ modes, and new physics effect in $B\to \phi
K_S$.
\end{abstract}


\section{Introduction}

Exclusive $B$ meson decays are important for extracting the
fundamental Standard Model parameters, such as the
Cabibbo-Kobayashi-Maskawa (CKM) matrix elements, and for exploring
new physics. They are complicated due to strong dynamics, which
must be well understood in order to achieve the above goals.
Especially, QCD theories for two-body nonleptonic decays are
necessary. Several approaches based on different factorization
theorems have been developed, which include perturbatiove QCD
(PQCD) \cite{LY1,KLS,LUY}, QCD-improved factorization (QCDF)
\cite{BBNS}, soft-collinear effective theory (SCET)
\cite{bfl,bfps}, light-cone sum rules (LCSR)
\cite{Chernyak:1990ag,sum-rules,Khodja}, and light-front QCD
(LFQCD) \cite{CZL,CJ99}. Most of them are based on collinear
factorization \cite{Ste}, but PQCD is based on $k_T$ factorization
\cite{CCH,CE,LRS}. In this talk I will briefly explain the
differences between collinear and $k_T$ factorizations, and review
recent progress on exclusive $B$ meson decays made in PQCD,
concentrating on the evolution of the $B$ meson wave function,
radiative decays, polarizations in $VV$ modes, and new physics
effect in $B\to \phi K_S$.

\section{Collinear vs. $k_T$ Factorization}

\begin{figure}[t!]
\begin{center}
\begin{picture}(100,100)
\Line(10,75)(90,75)
\Line(10,25)(90,25)
\CArc(10,65)(10,0,180)
\CArc(10,35)(10,-180,0)
\Line(0,35)(0,65)
\Line(20,35)(20,65)
\CArc(90,65)(10,0,180)
\CArc(90,35)(10,-180,0)
\Line(80,35)(80,65)
\Line(100,35)(100,65)
\Photon(60,75)(60,100){3}{3}
\Gluon(40,25)(40,75){5}{4}
\Text(10,50)[]{$\phi_\pi$}
\Text(90,50)[]{$\phi_\pi$}
\Text(50,0)[]{$\alpha_s$}
\end{picture} \hspace{0.5cm}
\begin{picture}(100,100)
\Line(10,75)(90,75)
\Line(10,25)(90,25)
\CArc(10,65)(10,0,180)
\CArc(10,35)(10,-180,0)
\Line(0,35)(0,65)
\Line(20,35)(20,65)
\CArc(90,65)(10,0,180)
\CArc(90,35)(10,-180,0)
\Line(80,35)(80,65)
\Line(100,35)(100,65)
\Photon(50,75)(50,100){3}{3}
\Gluon(35,25)(35,75){5}{4}
\Gluon(65,25)(65,75){5}{4}
\Text(10,50)[]{$\phi_\pi$}
\Text(90,50)[]{$\phi_\pi$}
\Text(50,0)[]{$\alpha_s^2$}
\end{picture} \hspace{0.5cm}
\begin{picture}(100,100)
\Line(10,75)(90,75)
\Line(10,25)(90,25)
\CArc(10,65)(10,0,180)
\CArc(10,35)(10,-180,0)
\Line(0,35)(0,65)
\Line(20,35)(20,65)
\CArc(90,65)(10,0,180)
\CArc(90,35)(10,-180,0)
\Line(80,35)(80,65)
\Line(100,35)(100,65)
\Photon(60,75)(60,100){3}{3}
\Gluon(40,25)(40,75){5}{4}
\Text(10,50)[]{$\phi_p$}
\Text(90,50)[]{$\phi_p$}
\Text(50,0)[]{$\alpha_s\Lambda^2/Q^2$}
\end{picture}
\end{center}
\caption{Perturbative expansion of $F_\pi$ in $\alpha_s$ and in
$\Lambda/Q$.}\label{fig1}
\end{figure}
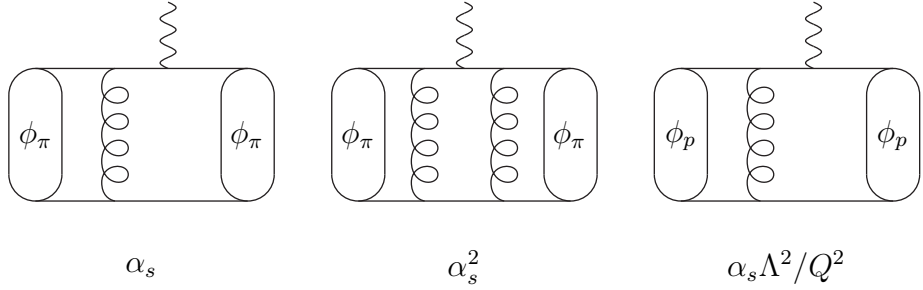

According to factorization theorems, the amplitude for an
exclusive process is calculated as an expansion of $\alpha_s(Q)$
and $\Lambda/Q$, where $Q$ denotes a large momentum transfer, and
$\Lambda$ is a small hadronic scale. For exclusive processes, such
as hadron form factors, collinear factorization was developed in
\cite{BL,ER,CZS,CZ}. Take the pion form factor $F_\pi$ involved in
the scattering process $\pi\gamma^*\to\pi$ as an example.
$F_{\pi}$ is expressed, up to next-to-leading order and
next-to-leading power (incomplete), as
\begin{eqnarray}
F_\pi=\phi_\pi\otimes H^{(0)}\otimes \phi_\pi
+\phi_\pi\otimes H^{(1)}\otimes \phi_\pi
+\phi_p\otimes H^{\prime (0)}\otimes \phi_p\;,
\end{eqnarray}
with each term being indicated in Fig.~\ref{fig1}. In the above
expression $\phi_\pi$ is the nonperturbative pion wave function,
$\phi_p$ the two-parton twist-3 pion wave function, and $H$ the
perturbative hard kernels. $\otimes$ stands for the convolutions
in parton momentum fractions in collinear factorization, and in
both parton momentum fractions and transverse momenta in $k_T$
factorization.

A parton momentum fraction $x$ must be integrated over in the
range between 0 and 1. Hence, the end-point region with a small
$x$ is not avoidable. If there is no end-point singularity
developed in a formula, collinear factorization works. If such a
singularity occurs, indicating the breakdown of collinear
factorization, $k_T$ factorization should be employed
\cite{BS,LS}. Moreover, factorization theorems do not only state
the separation of perturbative and nonperturbative dynamics, but
require controllable subleading effects.

Collinear factorization theorem for the semileptonic decay
$B\to\pi l\nu$ can be derived in a similar way \cite{L1}.
The involved $B\to\pi$ transition form factor $F^{B\pi}$
is expressed as
\begin{eqnarray}
F^{B\pi}=\int dx_1dx_2\phi_B(x_1)H(x_1,x_2) \phi_\pi(x_2)\;,
\end{eqnarray}
with the lowest-order hard kernel $H^{(0)}\propto
(1+2x_2)/(x_1x_2^2)$. The parton momentum fractions $x_1$ and
$x_2$ are carried by the spectator quarks
on the $B$ meson and pion sides, respectively. Obviously, the
above integral is logarithmically divergent for the asymptotic
model $\phi_\pi\propto x(1-x)$ \cite{SHB}.

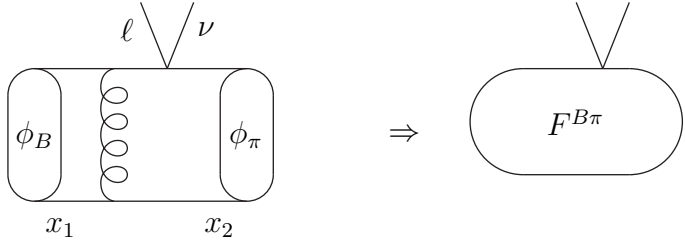
\begin{figure}[t!]
\begin{center}
\begin{picture}(100,100)
\Line(10,75)(90,75) \Line(10,25)(90,25) \CArc(10,65)(10,0,180)
\CArc(10,35)(10,-180,0) \Line(0,35)(0,65) \Line(20,35)(20,65)
\CArc(90,65)(10,0,180) \CArc(90,35)(10,-180,0) \Line(80,35)(80,65)
\Line(100,35)(100,65) \Line(50,100)(60,75) \Line(60,75)(70,100)
\Gluon(40,25)(40,75){5}{4} \Text(10,50)[]{$\phi_B$}
\Text(90,50)[]{$\phi_\pi$} \Text(45,90)[]{$\ell$}
\Text(75,90)[]{$\nu$} \Text(20,15)[]{$x_1$} \Text(80,15)[]{$x_2$}
\Text(150,50)[]{$\Rightarrow$}
\end{picture} \hspace{2cm}
\begin{picture}(100,100)
\Line(30,75)(70,75) \Line(30,35)(70,35) \CArc(30,55)(20,90,270)
\CArc(70,55)(20,-90,90) \Line(50,100)(60,75) \Line(60,75)(70,100)
\Text(50,55)[]{$F^{B\pi}$}
\end{picture}
\end{center}
\caption{$F^{B\pi}$ in $k_T$ factorization and in collinear
factorization.}\label{fig3}
\end{figure}

There are different ways to handle the end-point singularity:

An end-point singularity in collinear factorization implies that
exclusive $B$ meson decays are dominated by soft dynamics.
Therefore, a heavy-to-light form factor is not calculable, and
$F^{B\pi}$ should be treated as a soft object \cite{BBNS}. As
shown in Fig.~\ref{fig3}, it is meaningless to consider a
convolution of a hard kernel with meson wave functions, all of
which should be parameterized into a single nonperturbative
$F^{B\pi}$. This is the basis of QCDF, and subleading corrections
can be added systematically \cite{BF}.

The above treatment has been further elucidated in the framework
of SCET \cite{BPS}. In fact, only the 1 term in $H^{(0)}$ contains
the end-point singularity, which leads to an $O(\Lambda)$ object
$f^{\rm NF}$. The $2x_2$ term does not, leading to an
$O(\sqrt{m_B\Lambda})$ object $f^{\rm F}$. Therefore, at leading
power in $1/m_B$ and all orders in $\alpha_s$, the $B\to \pi$ form
factor can be split into the factorizable and nonfactorizable
components,
\begin{eqnarray}
F^{B\pi}(E) &=& f^{\rm F}(E) +f^{\rm NF}(E)\;,
\nonumber\\
f^{\rm NF}(E)&=&C(E,\mu_I)\zeta(\mu_I,E)\;,\nonumber\\
f^{\rm F}(E)&=&\phi_B(x_1,\mu) \otimes T'(E,\sqrt{E\Lambda},\mu)
\otimes \phi_{\pi}(x_2,\mu)\;, \label{fF}
\end{eqnarray}
with the factorization scales $\sqrt{E\Lambda}<\mu_I<E$ and
$\mu<\sqrt{E\Lambda}$. The kernel $T'$ has been further factorized
into a hard function characterized by the scale $m_B$ and a jet
function $J$ characterized by the scale $\sqrt{m_B\Lambda}$. As
shown above, the contributions characterized by $m_B$ and
$\sqrt{m_B\Lambda}$ have been clearly separated. The end-point
singularity arises only in the soft, nonperturbative form factors
$\zeta$, which obey the large-energy symmetry relations. $f^{\rm
NF}$ and $f^{\rm F}$ have been determined from the fit to the
$B\to\pi\pi$ data  recently \cite{BPRS}.

The third way is to adopt $k_T$ factorization theorem. When the
parton transverse momenta are included, $f^{\rm NF}$ does not
develop an end-point singularity, and both $f^{\rm NF}$ and
$f^{\rm F}$ are factorizable. $F^{B\pi}$ is then written as the
convolution \cite{LY1,TLS},
\begin{eqnarray}
F^{B\pi}=\int dx_1 dx_2 d^2k_{1T} d^2k_{2T}
\phi_B(x_1,k_{1T})H(x_1,x_2,k_{1T},k_{2T}) \phi_\pi(x_2,k_{2T})\;,
\end{eqnarray}
with the lowest-order hard kernel,
\begin{eqnarray}
H^{(0)}\propto \frac{1+2x_2}{[x_1x_2m_B^2+({\bf k}_{1T}-{\bf
k}_{2T})^2] [x_2m_B^2+k_{2T}^2]}\;,
\end{eqnarray}
as shown by the left-hand side of Fig.~\ref{fig3}. The end-point
singularity is smeared into the large logarithm $\ln^2(xm_B/k_T)$.
Resumming this logarithm to all orders in the conjugate $b$ space
\cite{Co03}, we have derived the Sudakov factor $S(xm_B, b)$,
which describes the parton distribution in $b$. Since $f^{\rm NF}$
has been included, the large-energy symmetry is respected in PQCD.
I mention that $k_T$ factorization theorem has been employed in
small-$x$ physics for decades.

\section{Ingredients of PQCD}

\subsection{Predictive power}

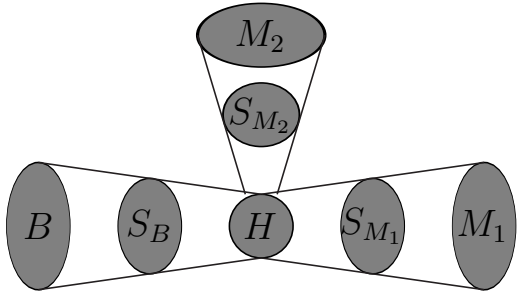
\begin{figure}[t!]
\begin{center}
\vspace{45pt} \hspace{20mm}\scalebox{1.2}{
\begin{picture}(90,0)(90,20)
\GOval(30,-10)(20,10)(0){0.5} \GOval(65,-10)(15,10)(0){0.5}
\GOval(135,-10)(15,10)(0){0.5} \GOval(170,-10)(20,10)(0){0.5}
\GOval(100,-10)(10,10)(0){0.5} \GOval(100,25)(10,12)(0){0.5}
\GOval(100,50)(10,20)(0){0.5} \Line(100,0)(30,10)
\Line(170,10)(100,0) \Line(80,50)(95,0) \Line(105,0)(120,50)
\Line(30,-30)(100,-20) \Line(100,-20)(170,-30)
\Text(65,-10)[]{$S_B$} \Text(30,-10)[]{$B$} \Text(100,-10)[]{$H$}
\Text(135,-10)[]{$S_{M_1}$} \Text(170,-10)[]{$M_1$}
\Text(100,25)[]{$S_{M_2}$} \Text(100,50)[]{$M_2$}
\end{picture}}
\end{center}
\vskip 2.5cm \caption{PQCD picture for two-body nonleptonic $B$
meson decays.}\label{fig4}
\end{figure}

The PQCD factorization picture of two-body nonleptonic $B$ meson
decays is shown in Fig.~\ref{fig4}. The parton transverse momenta
$k_T$, just coming out of or entering the mesons, are of
$O(\Lambda_{\rm QCD})$. With infinitely many collinear gluon
emissions, $k_T$ accumulate and reach the hard scale
$\sqrt{m_B\Lambda_{\rm QCD}}$, such that the hard kernel is free
of the end-point singularity. This effect is called Sudakov
suppression. The different transverse sizes of the initial-state
and final-state mesons and of the hard scattering have been made
explicit in Fig.~\ref{fig4}, and the evolution between these
different sizes is described by the Sudakov factors $S$.
Consequently, all topologies of diagrams for two-body decays are
calculable in PQCD, including nonfactorizable and annihilation
diagrams. PQCD can thus go beyond the naive factorization
assumption.

The only inputs in PQCD are meson distribution amplitudes, whose
information can be obtained from QCD sum rules and lattice QCD.
There are no free parameters, such as the end-point cutoffs
$\rho_H$, $\rho_A$, $\cdots$ in QCDF. Soft physics is under
control with Sudakov suppression (see Page 271 of \cite{GS} and
other independent investigations in \cite{WY,NM}). Therefore, PQCD
has more predictive power than other approaches.

\subsection{Penguin enhancement}

As stated above, factorizable amplitudes are characterized by the
scale of $O(\sqrt{m_B\Lambda})$. In PQCD, all amplitudes are
factorizable, implying that a decay mode, if dominated by penguin
contributions, will be enhanced by the large Wilson coefficients
$C_{3-6}(\sqrt{m_B\Lambda})$ significantly. The dynamical
enhancement of penguin contributions is one of the unique features
of the PQCD approach. In $B\to PP$ modes with the pseudoscalar
meson $P$ emitted from the weak vertex, the contributions from the
$(S-P)(S+P)$ penguin operators are proportional to $m_0$ and to
the Wilson coefficients $C_{5,6}$. Therefore, the penguin
contributions are enhanced chirally by the larger $m_0$ in QCDF
and dynamically by the larger Wilson coefficients in PQCD. The
predictions for $B\to PP$ branching ratios are then roughly equal
from the two approaches, and the penguin enhancing mechanism can
not be distinguished in these modes.

However, $B\to VP$, $VV$ modes with the vector meson $V$ emitted
from the weak vertex do not involve the chiral scale $m_0$, and
the chiral enhancement is absent in QCDF. They still depend on the
Wilson coefficients $C_{5,6}$, and exhibit the dynamical
enhancement in PQCD. Hence, the predictions for $B\to VP$, $VV$
branching ratios from the two approaches can differ by a factor 2,
$B({\rm PQCD})/B({\rm QCDF})\sim 2$. The predicted $B\to\phi K$
branching ratios from PQCD \cite{CKL,M,MS03} and from QCDF
\cite{CY,HMW,Du,BN03} are summarized in Table~\ref{tab0} for a
comparison. The PQCD results have been updated by adopting the
Ball and Boglione model \cite{BB} for the twist-2 kaon
distribution amplitude, which become a bit smaller $(10.2\to 9.3,
9.6\to 8.5)$ \cite{MS03}.

\begin{table}[ht]
\begin{center}
\begin{tabular}{c c c c}\hline \hline
Branching ratio&  Data (PDG)  & PQCD & QCDF\\
\hline $B(\phi K^\pm)$ ($10^{-6}$)&$9.3\pm
1.0$&$9.3^{+3.1}_{-2.1}$ &$
4.5^{+0.5+1.8+1.9+11.8}_{-0.4-1.7-2.1-3.3}$\\
$B(\phi K^0)$ ($10^{-6}$)&$8.6^{+1.3}_{-1.1}$&$ 8.5^{+3.0}_{-2.0}$
&$
4.1^{+0.4+1.7+1.8+10.6}_{-0.4-1.6-1.9-3.0}$\\
\hline \hline
\end{tabular}
\end{center}
\caption{Comparison of the predicted $B\to\phi K$ branching ratios
with the experimental data.}\label{tab0}
\end{table}

\subsection{Strong phases}

In $k_T$ factorization, a (short-distance) strong phase is
generated from an on-shell internal particle,
\begin{eqnarray}
\frac{1}{xm_B^2-k_{T}^2+i\epsilon} =\frac{P}{xm_B^2-k_{T}^2}
-i\pi\delta(xm_B^2-k_{T}^2)\;.
\end{eqnarray}
The power counting rules for topological amplitudes vary with the
factorization pictures of exclusive $B$ meson decays. Therefore,
the important sources of strong phases are different in QCDF and
in PQCD. Since the leading factorizable emission amplitude is
real, strong phases arise from nonfactorizable emission diagrams
and from annihilation diagrams. In QCDF the $O(\alpha_s)$ weak
vertex correction is the source of strong phases, which is a bit
larger than the $O(\alpha_s m_0/m_B)$ annihilation amplitude.
Being subleading in $\alpha_s$, a strong phase obtained from QCDF
is expected to be small. In PQCD the important source is the
annihilation amplitude, which is only suppressed slightly by a
power of $m_0/m_B\sim O(1)$. That is, a strong phase obtained from
PQCD will be larger.

\begin{table}[ht]
\begin{center}
\begin{tabular}{c c c c c}\hline \hline
Mode& Belle & Babar  & PQCD & QCDF\\
\hline $K^+\pi^-$ &$-0.088\pm 0.035\pm 0.013$
\cite{BelC}&$-0.133\pm 0.030\pm
0.009$ \cite{BarA}&$-0.129\sim -0.219$ & $+0.05\pm 0.09$\\
$\pi^+\pi^-$&$+0.58\pm 0.15\pm 0.07$ \cite{BelA}&$+0.30\pm 0.25\pm
0.04$ \cite{babar} &$ +0.16\sim +0.30$&$-0.06\pm 0.12$\\\hline
\hline
\end{tabular}
\end{center}
\caption{Comparison of the predicted $B^0\to K^+\pi^-$,
$\pi^+\pi^-$ direct CP asymmetries with the experimental
data.}\label{tab7}
\end{table}

\begin{figure}[t!]
\begin{center}
\epsfig{file=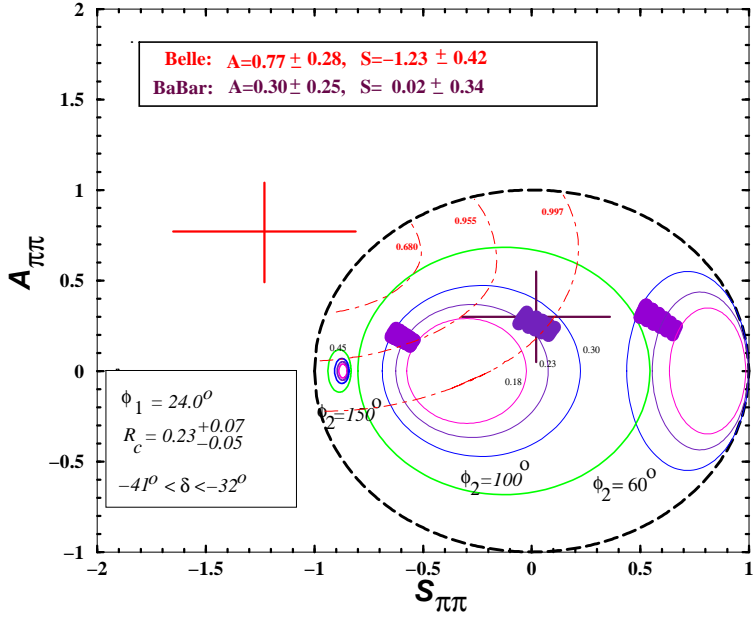,height=4.0in, angle=-90}
\caption{Correlation between $A_{\pi\pi}$ and $S_{\pi\pi}$
predicted by PQCD.}\label{fig5}
\end{center}

\end{figure}

Explicit calculations have shown that a strong phase from PQCD is
large and opposite in sign compared to that from QCDF. Because a
direct CP asymmetry $A_{CP}$ in $B$ meson decays is proportional
to the sine of the strong phase, the predictions from the two
approaches are very different \cite{KLS,LUY,Keum,Ben}:
$A_{CP}({\rm PQCD})/A_{CP}({\rm QCDF})\sim  -4$ as shown in
Table~\ref{tab7} for the $B^0\to K^+\pi^-$, $\pi^+\pi^-$ decays.
Hence, a comparison with experimental data can discriminate the
two approaches. It is observed that the central value of
$A_{\pi\pi}\sim 30\%$ measured by BaBar \cite{babar} is consistent
with the PQCD prediction. The predicted $S_{\pi\pi}$ varies with
the angle $\phi_2$. The central value of $S_{\pi\pi}\sim 0$ then
corresponds to $\phi_2\sim 80^o$ as shown in Fig.~\ref{fig5}.
Within the $90\%$ C.L. interval, the range of $\phi_2$,
$60^o<\phi_2<100^o$, has been extracted \cite{Keum}.

\section{Recent Progress}

\subsection{$B$ meson wave function}

The $B$ meson distribution amplitude, being the nonperturbative
input for collinear factorization, is defined via the nonlocal
matrix element,
\begin{eqnarray}\label{LCDA}
 & &  \langle\,0\,|\,\bar q(y^-)\,P\exp\left[-ig\int_0^{y^-}dz n_-\cdot
A(zn_-)\right]\, \Gamma\not n_- h(0)\,
    |\bar B(v)\rangle\nonumber\\
  & &\hspace{1.0cm}  = - \frac{if_B\sqrt{m_B}}{2}
  {\tilde \phi}_B(y^-,\mu)\,
    \mbox{tr}\left( \Gamma\not n_-\frac{1+\not v}{2}\,\gamma_5
    \right)\;,
\end{eqnarray}
where $n_-=(0,1,{\bf 0}_T)$ is a null vector, $h$ the rescaled $b$
quark field characterized by the $B$ meson velocity $v$, $\mu$ the
renormalization scale, and $\Gamma$ represents a Dirac matrix. The
property of the $B$ meson distribution amplitude has been studied
intensively in the literature \cite{GN,KKQT}.
Especially, the asymptotic behavior of $\phi_B$ has been extracted
from an evolution equation derived in the collinear factorization
theorem \cite{Neu03}.

\begin{figure}[t!]
\begin{center}
\begin{picture}(250,150)(-50,150)

\Line(-100,253)(-30,253) \Line(-100,250)(-30,250)
\Line(-100,185)(-30,185) \GlueArc(-30,250)(30,-180,-90){5}{4}
\Line(-30,220)(-30,250) \Line(-27,220)(-27,250)
\Text(-60,220)[]{$l$} \Text(-110,250)[]{$v$}
\Text(-17,220)[]{$n_-$} \Text(-60,160)[]{$(a)$}

\Line(0,253)(70,253) \Line(0,250)(70,250) \Line(0,185)(70,185)
\Gluon(40,250)(70,215){5}{4} \Line(70,185)(70,215)
\Line(73,185)(73,215) \Text(40,230)[]{$l$} \Text(40,160)[]{$(b)$}

\Line(100,253)(170,253) \Line(100,250)(170,250)
\Line(100,185)(170,185) \Gluon(140,185)(170,220){5}{4}
\Line(170,220)(170,250) \Line(173,220)(173,250)
\Text(140,160)[]{$(c)$}

\Line(200,253)(270,253) \Line(200,250)(270,250)
\Line(200,185)(270,185) \GlueArc(270,185)(30,90,180){5}{4}
\Line(270,185)(270,215) \Line(273,185)(273,215)
\Text(240,160)[]{$(d)$}

\end{picture}
\caption{Part of the $O(\alpha_s)$ diagrams for
$\phi_B$.}\label{fig6}
\end{center}
\end{figure}
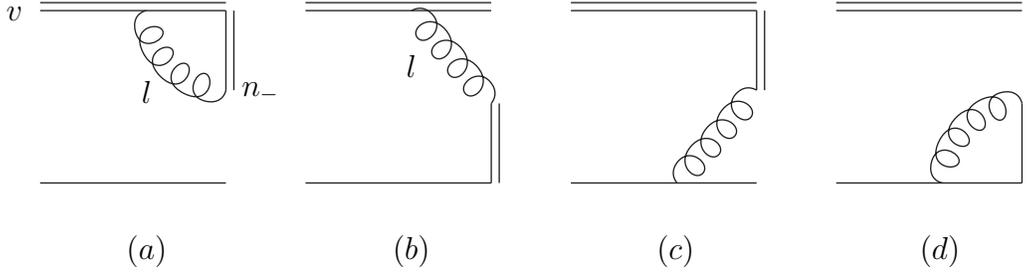

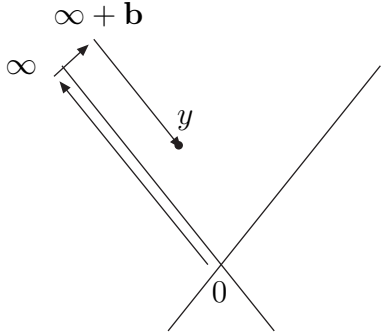
\begin{figure}[t!]
\begin{center}
\begin{picture}(300,150)(0,0)
\Line(130,20)(210,120) \Line(90,120)(170,20)
\LongArrow(145,45)(90,113) \Vertex(134,90){1.6}
\LongArrow(87,116)(99,127) \LongArrow(102,130)(134,90)
\Text(150,35)[]{{$0$}} \Text(75,120)[]{{$\infty$}}
\Text(104,140)[]{{$\infty+{\bf b}$}} \Text(137,100)[]{{$y$}}
\Text(50,35)[]{{\Large }}

\end{picture}
\caption{The path of $s$.}\label{fig7}
\end{center}
\end{figure}

The analysis starts with the evaluation of the $O(\alpha_s)$
diagrams in Fig.~\ref{fig7}, which are drawn according to
Eq.~(\ref{LCDA}). Figures~\ref{fig7}(a) and \ref{fig7}(b) give the
correction,
\begin{eqnarray}
Z^{(1)}_{ab}(k^+,k^{\prime +},\mu)&=&-ig^2C_F\mu^{2\epsilon}
\int\frac{d^{4-2\epsilon}l}{(2\pi)^{4-2\epsilon}} \frac{n_-\cdot
v}{v\cdot ll^2 n_-\cdot l}\nonumber\\
& &\times \left[\delta(k^+-k^{\prime +})-\delta\left(k^+-k^{\prime
+}+l^+\right)\right]\;.
\end{eqnarray}
The loop integral leads to the counterterm,
\begin{eqnarray}
-\frac{\alpha_sC_F}{2\pi}{ \frac{1}{\epsilon}}\left[
\frac{k^+}{k^{\prime +}}\frac{\theta(k^{\prime +}-k^+)}{(k^{\prime
+}-k^+)_+} +\frac{\theta(k^+-k^{\prime +})}{(k^+-k^{\prime
+})_+}\right]\;, \label{peB3}
\end{eqnarray}
where $1/\epsilon$ comes from the integration over $l_T$. The
corresponding anomalous dimension, the splitting kernel, then
determines the asymptotic behavior,
\begin{eqnarray}
{ \phi_B(k^+,\mu)\sim 1/k^+\;,\;\;{\rm for}\;\;k^+\to \infty}\;.
\end{eqnarray}
That is, the evolution effect ruins the normalizability of the $B$
meson distribution amplitude, and $f_B$ is not defined
unambiguously. This feature has been confirmed in a QCD sum rule
analysis \cite{BIK}.

Adopting $k_T$ factorization \cite{NL2,LL04}, Eq.~(\ref{LCDA}) is
modified into
\begin{eqnarray}
\langle 0|{\bar q}(y^-,{ b}) P\exp\left[-ig\int_0^{(y^-,{
b})}ds\cdot A(s)\right]\Gamma \not n_-h(0) |{\bar B}(v)\rangle\;,
\end{eqnarray}
where the path of $s$ for the Wilson line consists of three pieces
shown in Fig.~\ref{fig7}. The Feynman rule for Figs.~\ref{fig7}(a)
and \ref{fig7}(b) is modified into
\begin{eqnarray}
Z^{(1)}_{ab}(k^+,k^{\prime +},b,\mu) &=&-ig^2C_F\mu^{2\epsilon}
\int\frac{d^{4-2\epsilon}l}{(2\pi)^{4-2\epsilon}} \frac{n_-\cdot
v}{v\cdot ll^2 n_-\cdot l} \nonumber\\
& &\times \left[ \propto\delta(k^+-k^{\prime
+})-\delta\left(k^+-k^{\prime +}+l^+\right) { \exp(-i{\bf
l}_T\cdot {\bf b})} \right]\;.
\end{eqnarray}
The extra Fourier factor $\exp(-i{\bf l}_T\cdot {\bf b})$
suppresses the large $l_T$ region, such that the integral is UV
finite:
\begin{eqnarray}
-\frac{\alpha_sC_F}{\pi}\left[\frac{\theta(k^{\prime
+}-k^+)}{(k^{\prime +}-k^+)_+} K_0\left((k^{\prime
+}-k^+)b\right)-\frac{\theta(k^+-k^{\prime +})}{(k^+-k^{\prime
+})_+} K_0\left((k^+-k^{\prime +})b\right)\right]\;.
\end{eqnarray}
Therefore, The evolution effect does not involve a splitting
kernel, and does not change the $k^+$ distribution. The $B$ meson
distribution amplitude is regarded as the small $b$ limit, $b\to
1/m_B \to 0$, of the $B$ meson wave function, at which the Bessel
function $K_0$ remains UV finite. The above observation indicates
that $k_T$ factorization is a more appropriate tool for exclusive
$B$ meson decays than collinear factorization.

\subsection{Radiative decays}

The PQCD approach has been applied to radiative $B$ meson decays
recently. We first summarize the experimental data for the $B\to
K^*\gamma$ decays and the PQCD predictions \cite{KMS} in
Tables.~\ref{tab1} and \ref{tab2}, respectively, where the isospin
breaking is defined as
\begin{eqnarray}
\Delta_{0+} &\equiv & \frac{\Gamma(B^0\to
K^{*0}\gamma)-\Gamma(B^+\to K^{*+}\gamma)} {\Gamma(B^0\to
K^{*0}\gamma)+\Gamma(B^+\to K^{*+}\gamma)}\;.
\end{eqnarray}
Note that the large $B\to K^*\gamma$ branching ratios from NLO
QCDF \cite{BVgam} just imply that the input form factor value
$F^{BK^*}$ is too large \cite{A02,CC04}.

\begin{table}[ht]
\begin{center}
\vspace*{0.5cm}
\begin{tabular}{c c c }\hline \hline
Quantities& Belle \cite{Nakao:2004th} & Babar \cite{Aubert:2001me} \\
\hline $B(K^{*0} \gamma)(10^{-5})$&$4.01\pm 0.21\pm 0.17
$&$3.92\pm 0.20\pm 0.24$\\
$B(K^{*\pm} \gamma)(10^{-5})$&$4.25\pm 0.31\pm 0.24$&$3.87\pm
0.28\pm 0.26$\\
$\mbox{A}_{CP}$&$-0.015\pm 0.044\pm 0.012$&$-0.013 \pm 0.036\pm
0.010$
\\
$\Delta_{0+}$&$+0.012\pm 0.044\pm 0.026$&$+0.051\pm 0.044 \pm
0.023 \pm 0.024$\\\hline \hline
\end{tabular}
\end{center}
\caption{Data of the $B\to K^*\gamma$ decays.}\label{tab1}
\end{table}

\begin{table}[ht]
\begin{center}
\vspace*{0.5cm}
\begin{tabular}{c c c}\hline \hline
Quantities & PQCD & QCDF\\
\hline $B(K^{*0} \gamma)(10^{-5})$
&$3.5^{+1.1}_{-0.8}$& { 7.4} ($\mu=m_b$)\\
$B(K^{*\pm} \gamma)(10^{-5})$&$3.4^{+1.2}_{-0.9}$&\\
$\mbox{A}_{CP}$
&$(0.39^{+0.06}_{-0.07})\%$&$0.3\%$\\
$\Delta_{0+}$&${ +0.057^{+0.011}_{-0.013}}$&+0.038\\\hline \hline
\end{tabular}
\end{center}
\caption{PQCD predictions for the $B\to K^*\gamma$
decays.}\label{tab2}
\end{table}

\begin{figure}
\begin{center}
\includegraphics[width=5.0cm]{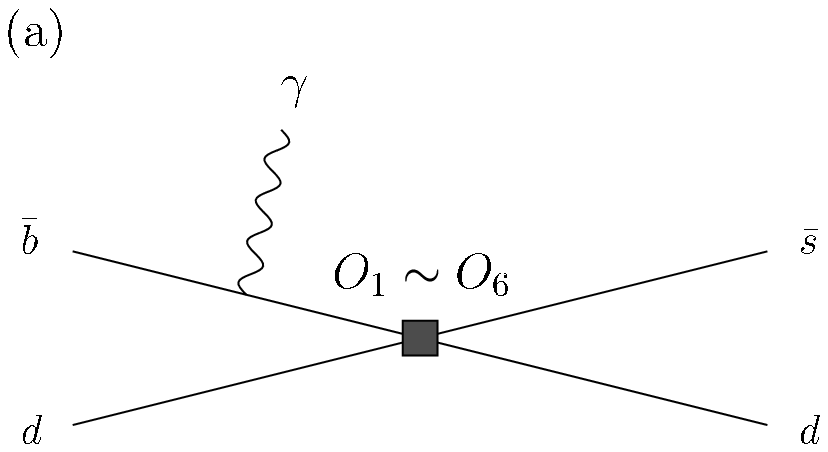}
\hspace{8mm}
\includegraphics[width=5.0cm]{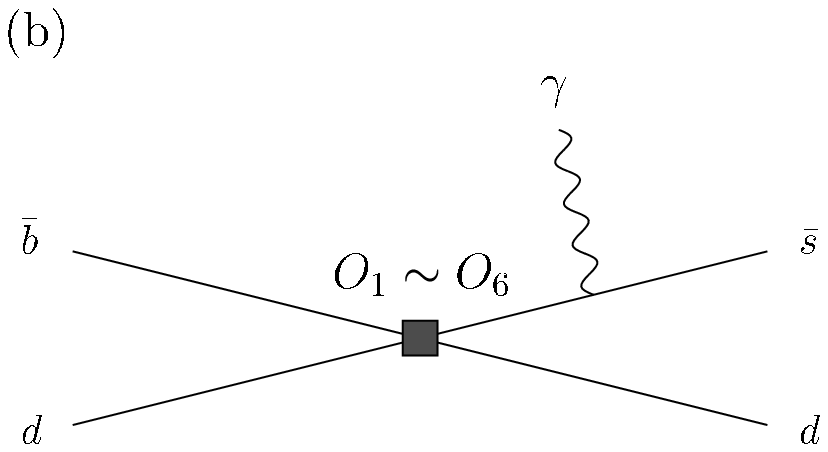}

\includegraphics[width=5.0cm]{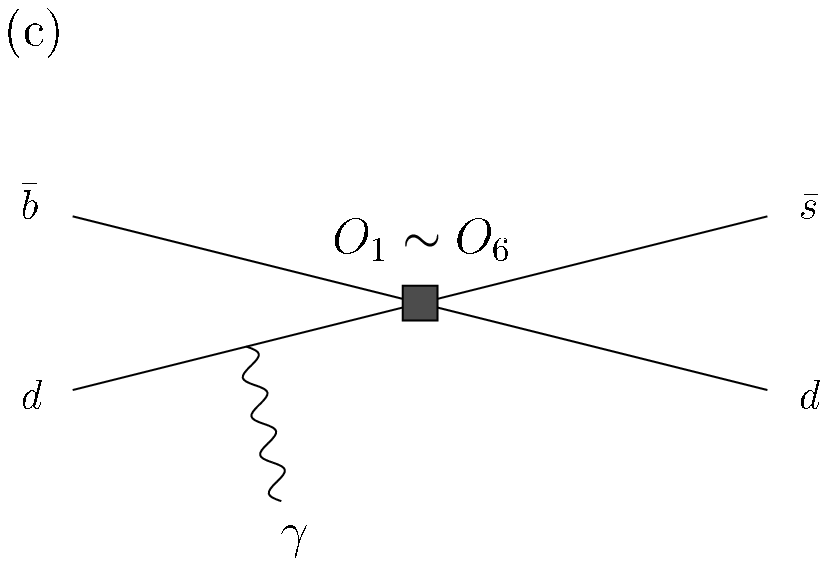}
\hspace{8mm}
\includegraphics[width=5.0cm]{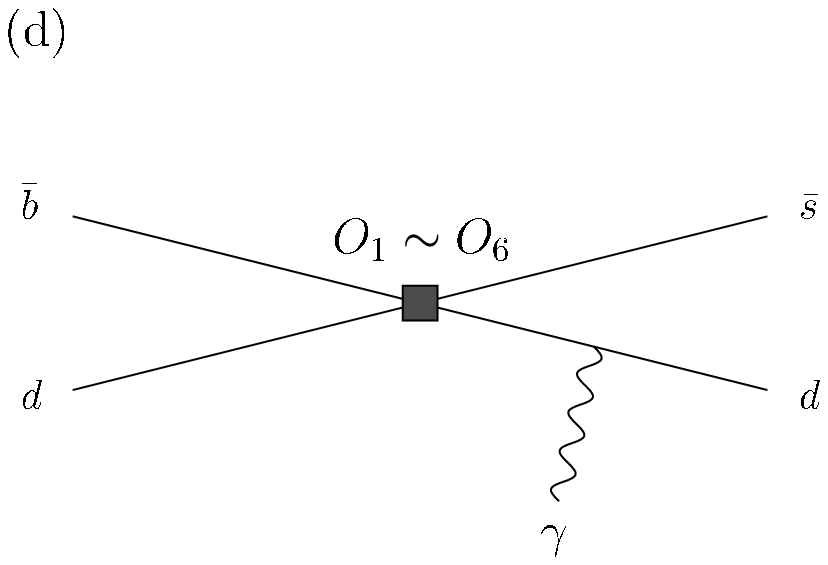}
\caption{Annihilation diagrams from ${O_{1}\sim O_6}$.}
\label{annihilation}
\end{center}
\end{figure}

After determining the $B$ meson wave function $\phi_B$ from the
$B\to\pi$ form factor, we can predict the above quantities in
PQCD. We have considered the contributions from the operators
${O_{7\gamma}}$, ${O_{8g}}$, and $O_2$ (up and charm loops) and
from the annihilation topology, and the long-distance contribution
through the vector mesons. The theoretical errors for the
branching ratios mainly come from the shape parameter in $\phi_B$,
and those for the CP asymmetry mainly from the CKM parameters. It
was found that the penguin annihilation from $O_{5,6}$ shown in
Fig.~\ref{annihilation} is the most important mechanism
responsible for the positive isospin breaking.

We have also analyzed the $B\to \rho (\omega)\gamma$ decays.
Preliminary PQCD predictions are listed in Table~\ref{tab3}, and
compared to those from QCDF. The small deviation between PQCDI and
PQCDII is attributed to the different treatment of the
long-distance contribution.

\begin{table}[ht]
\begin{center}
\vspace*{0.5cm}
\begin{tabular}{c c c c c}\hline \hline
Branching ratio& Belle \cite{Bel04} &PQCDI \cite{LMSY} &PQCDII \cite{AK} & QCDF \cite{B02} \\
\hline
$B(\rho^{\pm}\gamma)$ ($10^{-6}$)&$1.8^{+0.8}_{-0.7}\pm 0.1$&2.1&$1.8^{+0.4}_{-0.5}$& 1.6\\
$B(\rho^0\gamma)$ ($10^{-6}$)&$0.5^{+0.5}_{-0.4}\pm 0.2$&$0.95$&$0.9^{+0.3}_{-0.2}$&0.95\\
$B(\omega\gamma)$ ($10^{-6}$)&$1.3^{+0.7}_{-0.6}\pm 0.2$&$0.85$&&\\
\hline \hline
\end{tabular}
\end{center}
\caption{Branching ratios of the $B\to \rho (\omega)\gamma$
decays.}\label{tab3}
\end{table}

\subsection{Polarizations in $VV$ modes}

\begin{table}[ht]
\begin{center}
\vspace*{0.5cm}
\begin{tabular}{c  c c }\hline \hline
Pol. fraction&Belle \cite{belle,Zhang:2003up} &BaBar
\cite{Aubert:2003mm,Aubert:2003xc,gritsan}\\
\hline
$R_L(\phi K^{*+})$& &$0.46\pm0.12\pm0.03$\\
$R_L(\phi K^{*0})$&$0.43\pm0.09\pm0.04$&$0.52\pm0.07\pm0.02$\\
$R_\perp(\phi K^{*0})$&$0.41\pm0.10\pm0.02$&$0.27\pm0.07\pm0.02$\\
\hline
$R_L(\rho^0 K^{*+})$&&$0.96^{+0.04}_{-0.15}\pm0.04$\\
$R_L(\rho^0 \rho^+)$&$0.95\pm0.11\pm0.02$&$0.97^{+0.03}_{-0.07}\pm0.04$\\
$R_L(\rho^+ \rho^-)$&&$0.98^{+0.02}_{-0.08}\pm0.03$\\
\hline \hline
\end{tabular}
\end{center}
\caption{Polarization fractions in $B\to VV$
transitions.}\label{tab4}
\end{table}

The measured polarization fractions in $B\to VV$ modes are
summarized in Table~\ref{tab4}, whose implication will be
discussed below. The power counting rules for the emission
topologies are
\begin{eqnarray}
R_L\sim 1\;,\;\;\;\;R_\parallel \sim R_\perp\sim
O(m_\phi^2/m_B^2)\;,
\end{eqnarray}
and those for the annihilation topologies from $O_{5,6}$ are
\begin{eqnarray}
{ R_L\sim R_\parallel \sim R_\perp\sim O(m_{K^*}^2/m_B^2,
m_\phi^2/m_B^2)}\;.
\end{eqnarray}
If the emission topologies dominate as expected, the $B\to\phi
K^*$ data do not obey the counting rules obviously. If the
annihilation contribution is enhanced by some mechanism, one could
have $R_L\sim 0.5$. This is the strategy adopted in \cite{K04}
based on QCDF.

I stress that any proposed mechanism has to explain all $VV$ modes
simultaneously, especially $B\to \rho K^*$ \cite{CDP}, and that
$B\to\phi K^*$ is so unique, since all other modes follow the
expected counting rules. The annihilation amplitude is a free
parameter in QCDF, since it involves the arbitrary end-point
cutoff $\rho_A$. Because varying the free parameter to explain the
data can not be conclusive, it is better to estimate the
annihilation contribution in a reliable way. Viewing that the PQCD
predictions for the annihilation amplitudes are consistent with
the measured direct CP asymmetries in $B\to K\pi$, $\pi\pi$
\cite{KLS,LUY}, this calculation has been performed in
\cite{CKL2}. As shown in Table~\ref{tab5}, the annihilation
mechanism indeed helps, but is not sufficient to lower the
fraction $R_L$ down to around 0.5.

A mechanism for enhancing the transverse polarization component in
the $B\to \phi (\omega)K^*$ decays has been proposed in \cite{HN},
which arises from the $b\to sg$ transition. The novelty is that
the transverse polarization of the gluon from the transition
propagates into the $\phi (\omega)$ meson, and that the constraint
from the $B\to\rho K^*$ data is avoided. The relevant matrix
element was then parameterized in terms of a dimensionless free
parameter $\kappa$. By varying this parameter to $\kappa=-0.25$,
the authors of \cite{HN} claimed that the $B\to\phi K^*$ data
could be accommodated in the Standard Model. Our comment is that a
reliable estimate of the $\kappa$ value is necessary. By means of
the three-parton $\phi$ meson distribution amplitude and the naive
factorization assumption, we have found that the order of
magnitude of $\kappa$ is, unfortunately, 0.01. The detail will be
supplied elsewhere.

\begin{table}[htbp]
\begin{center}
\begin{tabular}{cccccc}
\hline
Mode & $ |A_{0}|^{2}$ & $ |A_{\parallel}|^{2}$ & $
|A_{\perp}|^{2}$ & $\phi_{\parallel}(rad.)$ & $\phi_{\perp}(rad.)$
\\ \hline
$\phi K^{*0}$(I) & $0.923$  & $0.040$ & $0.035$ &
$\pi$ & $\pi$\\
\hspace{0.7cm}(II)&  $0.860$ & $0.072$ & $0.063$ & $3.30$
& $3.33$ \\
\hspace{0.7cm}(III) &  $0.833$ & $0.089$ & $0.078$ &
$2.37$ & $2.34$ \\
\hspace{0.7cm}(IV) & { $0.750$} & $0.135$ & $0.115$ & $2.55$ & $2.54$ \\
\hline $\phi K^{*+}$(I)  &  $0.923$ & $0.040$ & $0.035$ &
$\pi$  & $\pi$  \\
\hspace{0.7cm}(II)  &$0.860$ & $0.072$ & $0.063$ & $3.30$ & $3.33$
\\
\hspace{0.7cm}(III) &$0.830$ & $0.094$ & $0.075$ &
$2.37$ & $2.34$ \\
\hspace{0.7cm}(IV)  &{ $0.748$} & $0.133$ & $0.111$ & $2.55$ & $2.54$ \\
\hline
\end{tabular}
\end{center}
\caption{(I) Without nonfactorizable and annihilation
contributions, (II) add  only nonfactorizable contribution, (III)
add only annihilation contribution, (IV) add both nonfactorizable
and annihilation contributions.}\label{tab5}
\end{table}

\subsection{New physics in $B\to \phi K_S$}

It has been known that the discrepancy between the induced CP
asymmetries $S_{J/\psi K_S}$ and $S_{\phi K_S}$ measured by Belle
could be a possible new physics signal. The data of the $B\to\phi
K_s$ mode are summarized below \cite{Aubert:2002ic,Abe:2003yu}:
\begin{eqnarray*}
S_{J/\psi K_S} &=& \left\{
\begin{array}{cl}
 0.741 \pm 0.067\pm 0.034 & (\mathrm{BaBar,\ 81\, fb}^{-1})\;,\\
 0.733\pm 0.057\pm 0.028 & (\mathrm{Belle,\ 140\, fb}^{-1})\;,\\
\end{array}
\right.\\
S_{\phi K_S} &=& \left\{
\begin{array}{cl}
 0.45\pm 0.43\pm 0.07 & \mathrm{(BaBar,\ 110\, fb}^{-1})\;,\\
 { -0.96\pm 0.50\ {}^{+0.09}_{-0.11}}
  & \mathrm{(Belle,\ 140\, fb}^{-1})\;.\\
\end{array}
\right.
\end{eqnarray*}
There are already many works devoted to the new physics study in
the above mode. Here I will not concentrate on the new physics
effect, but on the advantage of exploring new physics in the PQCD
framework.

It is expected that the new CP violating sources and the flavor
changing neutral current in MSSM induce direct CP violations, and
render $S_{\phi K_S}$ different from $S_{J/\psi K_S}$. Motivated
by this observation, we have calculated the Wilson coefficients
using the mass insertion approximation \cite{MS03}. For example,
the coefficient associated with the magnetic penguin operator is
given by
\begin{eqnarray}
C_{8G}^{\rm NP} (M_S) \simeq \frac{\sqrt{2} \alpha_s \pi} {2G_F
V_{ts}^* V_{tb} m_{\tilde{q}}^2} \left[ (\delta_{LL}^d)_{23}\left(
\frac{1}{3} M_3(x) + \!3 M_4(x)\right) +
(\delta_{LR}^d)_{23}\frac{m_{\tilde{g}}}{m_b} \left(\frac{1}{3}
M_1(x) + 3 M_2(x)\right)\right] \label{eq:cmp} \;,
\end{eqnarray}
where $M_S$ is the SUSY scale, $x =
m^2_{\tilde{g}}/m^2_{\tilde{q}}$ with $m_{\tilde g}$ and
$m_{\tilde q}$ being the gluino and squark masses, respectively,
and $B(x)$, $P(x)$ and $M(x)$ the loop functions from box and
penguin diagrams \cite{Gabbiani:1996hi,Harnik:2002vs}. The
relevant matrix element associated with $O_{8G}$ was then
calculated in PQCD. Note that such a calculation is ambiguous in
naive factorization because of the unknown gluon invariant mass
$q^2$. In PQCD, it is written as $q^2=(1-x_2)x_3M_B^2-|{\bf
k}_{2T}-{\bf k}_{3T}|^2$ with $x_2$ and $k_{2T}$ ($x_3$ and
$k_{3T}$) being the momentum fraction and the transverse momentum
in the $K$ ($\phi$) meson.

\begin{figure}[bt]
\includegraphics[width=17.0cm]{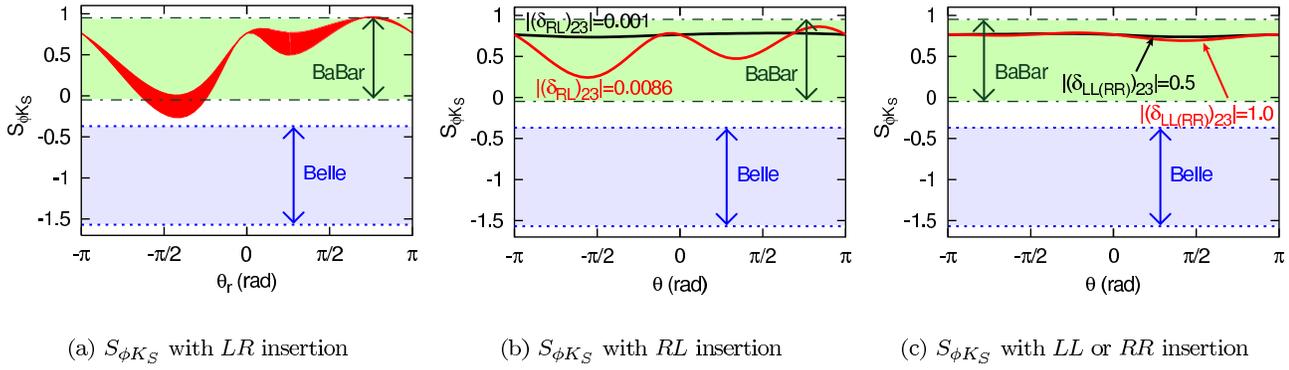}%
\caption{\label{phiks}The result of $S_{\phi K_S}$ with the MSSM
contribution. The dot-dashed (dotted) lines represent the
$1\sigma$ error from BaBar (Belle).}
\end{figure}

Next step is to constrain the parameters $(\delta_{LL}^d)_{23}$
$\cdots$ from the data of the branching ratio $B(B\to X_s \gamma)$
and of the $B_s-\overline{B_s}$ mixing, assuming $m_{\tilde
g}=m_{\tilde q}=500$ GeV. The predicted range of $S_{\phi K_S}$
\cite{MS03} is displayed in Fig.~\ref{phiks}. One can further
constrain the parameters from the $B\to\phi K$ branching ratios
and direct CP asymmetries. Due to the larger strong phases, and
the larger direct CP asymmetries from PQCD, one can extract a
stronger constraint on new physics from the data. Conservatively,
we have $S_{\phi K_S} \geq -0.28$ as exhibited in
Fig.~\ref{phiks}. Varying $m_{\tilde g}$ and $m_{\tilde q}$
arbitrarily, $S_{\phi K_S}$ reaches about $-40\%$. Therefore, it
is difficult to explain the Belle data by the considered new
physics.

\section{Summary}

In this talk I have summarized the recent progress on exclusive
$B$ meson decays made in the PQCD approach based on $k_T$
factorization theorem. Both collinear and $k_T$ factorization
theorems can be developed for these decays. In the former a
heavy-to-light transition form factor exhibits an end-point
singularity, while in the latter it is infrared-finite. Hence,
soft dominance is postulated and a heavy-to-light form factor is
parameterized as a nonperturbative input in collinear
factorization. Hard dominance is postulated and a heavy-to-light
form factor can be factorized into a convolution of a hard kernel
with meson wave functions in $k_T$ factorization. Note that a
mixed picture, in which soft and hard contributions are postulated
to be of the same order of magnitude, has been proposed in SCET
\cite{BPS}. As explained above, there is no conflict between the
soft-dominance and hard-dominance pictures for exclusive $B$ meson
decays, which are due to the different theoretical frameworks. The
$B\to\pi$ form factor $F^{B\pi}$ is not factorizable, partially
factorizable, and completely factorizable in QCDF, SCET, and PQCD,
respectively.

It has been found that the evolution of the $B$ meson distribution
amplitude in collinear factorization ruins its normalizability,
while the evolution of the $B$ meson wave function in $k_T$
factorization is well-behaved. The branching ratios, CP
asymmetries, and isospin breaking in radiative decays have been
calculated. The results are consistent with the data. The
annihilation contribution can be estimated in PQCD reliably. Its
effect helps, but is not sufficient for explaining the observed
$B\to\phi K^*$ polarization fractions. We have taken the induced
CP asymmetry in the $B\to\phi K_S$ mode as an example to
demonstrate that PQCD gives a stronger constraint on new physics
\cite{NK}.

I did not cover the following subjects in this talk: the
evaluation of the nonfactorizable contribution to the decays $B\to
D^{(*)}\pi(\rho,\omega)$ \cite{KKLL}, the analysis of three-body
decays by means of two-meson distribution amplitudes \cite{CL04},
and the studies of decays into scalar mesons \cite{Chen}.

\vskip 1.0cm I thank the KTPQCD group members for useful comments
and discussions. This work was supported in part by the National
Science Council of R.O.C. under Grant No. NSC-92-2112-M-001-030
and by Taipei branch of the National Center for Theoretical
Sciences of R.O.C..

\end{document}